\documentclass[intlimits,twoside,a4paper]{article}
\usepackage[utf8]{inputenc}
\usepackage[eqsecnum]{cmpj3}
\pdfoutput=1
\usepackage{graphics}
\usepackage{bm}
\newcommand{\be}{\begin{equation}}
\newcommand{\ee}{\end{equation}}
\newcommand{\beqn}{\begin{eqnarray}}
\newcommand{\eeqn}{\end{eqnarray}}

\definecolor{mymagenta}{rgb}{1.0,0.0,1.0}
\definecolor{mycyan}{rgb}{0.0,1.0,1.0}
\definecolor{myyellow}{rgb}{1.0,1.0,0.0}
\definecolor{myorange}{rgb}{1.0,0.27,0.0}

\definecolor{dark-gray}{HTML}{a0a0a0}
\definecolor{dark-red}{HTML}{8b0000}
\definecolor{dark-green}{HTML}{006400}
\definecolor{dark-blue}{HTML}{00008b}
\definecolor{gold}{rgb}{1.0,0.84,0.0}
\definecolor{dark-turquoise}{HTML}{00ced1}

\issue{2023}{26}{1}{13102}
\doinumber{10.5488/CMP.26.13102}

\title[Random Ising chain in transverse and longitudinal fields]{Random Ising chain in transverse and longitudinal fields:
Strong disorder RG study}
\author[T. Pet\H {o}, F. Igl{\'o}i, I. A. Kov\'acs]{T. Pet\H {o}\refaddr{label1},
        F. Igl{\'o}i\orcid{0000-0001-6600-7713}\refaddr{label2,label1}, I. A. Kov\'acs\orcid{0000-0002-6890-6277}\refaddr{label3,label4}\thanks{Corresponding author: \email{istvan.kovacs@northwestern.edu}.}}
\addresses{
\addr{label1} Institute of Theoretical Physics, Szeged University, H-6720 Szeged, Hungary
\addr{label2} Wigner Research Centre for Physics, Institute for Solid State Physics and Optics, H-1525 Budapest, P.O. Box 49, Hungary
\addr{label3} Department of Physics and Astronomy, Northwestern University, Evanston, IL 60208, USA
\addr{label4} Northwestern Institute on Complex Systems, Northwestern University, Evanston, IL 60208, USA
}


\bibstyle{cmpj}

\date{Received October 18, 2022, in final form November 21, 2022}

\begin{document}

\maketitle

\begin{abstract}
Motivated by the compound ${\rm LiHo}_x{\rm Y}_{1-x}{\rm F}_4$, we consider the Ising chain with random couplings and in the presence of simultaneous random transverse and longitudinal fields, and study its low-energy properties at zero temperature by the strong disorder renormalization group approach. In the absence of longitudinal fields, the system exhibits a quantum-ordered and a quantum-disordered phase separated by a critical point of infinite disorder. When the longitudinal random field is switched on, the ordered phase vanishes and the trajectories of the renormalization group are attracted to two disordered fixed points: one is characteristic of the classical random field Ising chain, the other describes the quantum disordered phase. The two disordered phases are separated by a separatrix that starts at the infinite disorder fixed point and near which there are strong quantum fluctuations.
%\end{abstract}
%
\printkeywords
%
%\pacs{}
\end{abstract}

\section{Introduction}
\label{sec:intr}
Quantum fluctuations play an important role in characterising the physical properties of systems in the vicinity of a quantum phase-transition point, which takes place at zero temperature by varying a control parameter, such as the strength of a transverse field \cite{sachdev_2011}. In the presence of quenched disorder, which is connected to the energy density, such as having random (ferromagnetic) couplings and/or transverse fields, the combined action of quantum and disorder fluctuations generally results in a new type of a phase transition. This is often controlled by a so-called infinite disorder fixed point (IDFP) \cite{FISHER1999222}. The prototype of such systems is the random transverse-field Ising (RTFI) chain with nearest neighbour interactions, the low-energy properties of which was calculated asymptotically exactly \cite{PhysRevLett.69.534,PhysRevB.51.6411} by a so-called strong disorder renormalization group (SDRG) method \cite{PhysRevLett.43.1434,PhysRevB.22.1305}, for reviews see \cite{IGLOI2005277,Igloi2018}. It was shown that at the IDFP, the renormalized couplings and transverse fields have an extremely broad distribution, so that the ratio of renormalized parameters at the neighbouring sites is either zero or infinity. Thus, the critical properties are fully dominated by disorder fluctuations. Using an efficient numerical implementation of the SDRG technique \cite{Kov_cs_2011}, the RTFI model was studied in higher dimensions \cite{PhysRevB.61.1160,10.1143/PTPS.138.479,Karevski2001,PhysRevLett.99.147202,PhysRevB.77.140402,PhysRevB.80.214416,PhysRevB.82.054437,PhysRevB.83.174207} and IDFP behaviour is observed in these cases, too. On the contrary, the RTFI model with long-range interactions has a conventional random critical behaviour \cite{Juh_sz_2014,PhysRevB.93.184203}.

An experimental realization of the RTFI model is the compound ${\rm LiHo}_x{\rm Y}_{1-x}{\rm F}_4$, which is a dipole coupled random Ising ferromagnet \cite{PhysRevB.42.4631,PhysRevLett.67.2076,PhysRevLett.71.1919,doi:10.1126/science.284.5415.779,dutta_aeppli_chakrabarti_divakaran_rosenbaum_sen_2015}. If this system is placed in a magnetic field, which is transverse to the Ising axis, it acts as an effective transverse field, which, however, also induces a random longitudinal field \cite{PhysRevLett.97.237203,PhysRevB.77.020401,Schechter_2009}. This experimental example motivates our study in this paper about the properties of the Ising model with random couplings and in the presence of simultaneous random transverse and longitudinal fields. Here, we consider the one-dimensional problem with nearest neighbour interactions and use an SDRG approach, in which the well known decimation rules for the random couplings and the random transverse fields \cite{IGLOI2005277,Igloi2018} are extended to those for the random longitudinal fields \cite{PhysRevB.101.024203,PhysRevB.104.174206}. We apply an efficient numerical algorithm and study the RG phase-diagram from the properties of the RG flow. In the vicinity of the IDFP, we also investigate the scaling behaviour of the excitation energy and the magnetization moment for large finite samples.

The rest of the paper is organised in the following way. The model is introduced in section~\ref{sec:model} and then in section~\ref{sec:SDRG} the SDRG method is explained. Numerical results for finite random longitudinal fields are presented in section~\ref{sec:results} and discussed in section~\ref{sec:disc}.

\section{The model}
\label{sec:model}
We consider the Ising chain with random couplings and in the presence of simultaneous random transverse and longitudinal fields, defined by the Hamiltonian:
\be
\hat{H}=-\sum_{i=1}^L J_i \sigma_{i}^{z} \sigma_{i+1}^{z}-\sum_{i=1}^L \Gamma_i \sigma^x_{i} -\sum_{i=1}^L h_i \sigma^z_{i}\;,
\label{Hamilton}
\ee
Here, the $\sigma_{i}^{x,z}$ are Pauli matrices at site $i$ and we use periodic boundary conditions: $\sigma_{L+1}^{z}\equiv \sigma_{1}^{z}$. The nearest neighbour couplings are ferromagnetic, $J_i>0$ and random, the transverse fields  $\Gamma_i>0$ are random, the longitudinal fields $h_i$ are random and are taken from a symmetric distribution: $p(h)=p(-h)$, so that there is no induced global magnetization in the longitudinal direction. %In this paper 
In the numerical work, we generally used the following box-like distributions:
\begin{align}
 % \begin{split}
    \pi_1(J) &=
    \begin{cases}
      1, & \hspace*{0.65cm}\text{for } 0<J\leqslant 1,\nonumber\\
      0, & \hspace*{0.65cm}\text{otherwise,}
    \end{cases} \\
    \pi_2(\Gamma) &= 
    \begin{cases}
      1/\Gamma_0,& \text{for } 0<\Gamma \leqslant \Gamma_0,\nonumber\\
      0, & \text{otherwise,} 
    \end{cases}\\
    p(h) &= 
    \begin{cases}
      1/h_0,& \text{for } -h_0/2 \leqslant h \leqslant h_0/2,\\
      0, & \text{otherwise.} 
    \end{cases}
 % \end{split} 
  \label{eq:J_distrib} 
\end{align}
Note that the non-random model, with $J_i=J$ and $\Gamma_i=\Gamma$ in the presence of a staggered longitudinal field $h_i=h(-1)^i$, is equivalent to the antiferromagnetic Ising model in transverse and longitudinal fields. This model was studied theoretically in references~\cite{PhysRevE.63.016112,PhysRevB.68.214406,PhysRevE.99.012122,PhysRevB.103.174404} and experimentally in reference~\cite{Simon2011}. For random couplings and random transverse fields, but with non-random staggered longitudinal fields, the model is studied in references~\cite{PhysRevB.96.064427,PhysRevB.101.024203}. Variations of the model were also studied in the context of many-body localization  \cite{MBL1,MBL2,MBL3,MBL4,MBL5,MBL6,MBL7}. 

\section{SDRG calculations}
\label{sec:SDRG}

In the SDRG procedure \cite{IGLOI2005277,Igloi2018}, we consider the separated local degrees of freedom, say at position $i$. These are couplings or sites having the value of the characteristic parameters: $J_i$ and 
\be
\gamma_i=\sqrt{\Gamma_i^2+h_i^2},
\label{epsilon_i}
\ee 
respectively. These parameters and the related gaps are sorted in descending order and the largest one, denoted by $\Omega$, which sets the energy-scale in the problem, is eliminated and between the remaining degrees of freedom new terms in the Hamiltonian are generated through perturbation calculation. This procedure is successively iterated, during which $\Omega$ is monotonously decreasing. At the fixed point, with $\Omega^*=0$
one makes an analysis of the distribution of the different parameters and calculates the scaling properties. In what follows, we describe the elementary decimation steps.

%\subsection{Elementary decimation steps}
%\label{sec:decimation}

%%%%%%%%%%%%%%%%%%%%%%%%% Fig 0 %%%%%%%%%%
\begin{figure}[h!]
\center
\includegraphics[width=.8\columnwidth]{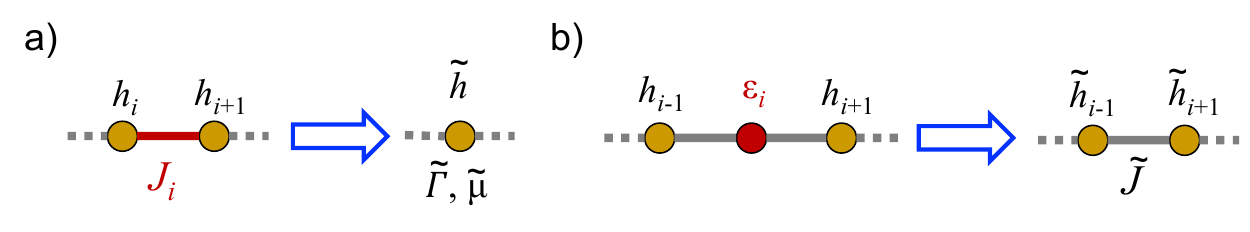}
	\vskip -0.4cm
\caption{(Colour online) Illustration of the SDRG decimation steps for strong-coupling (left-hand panel) and strong field (right-hand panel) decimation.}
\label{fig_0}	
\end{figure} 
%%%%%%%%%%%%%%%%%%%%%

\subsubsection{Strong-coupling decimation}

In this case, the largest local term in the Hamiltonian is a coupling, say $\Omega=J_i$, connecting sites $i$ and $i+1$ as illustrated in figure~\ref{fig_0}, and  the two-site Hamiltonian is given by:
\begin{equation}
\hat{H}_{cp}=-J_i \sigma^z_{i}\sigma^z_{i+1}-\Gamma_i\sigma^x_{i} -\Gamma_{i+1}\sigma^x_{i+1} - \left(h_i \sigma^z_{i}+h_{i+1} \sigma^z_{i+1}\right).
\label{strong-coupling}
\end{equation}
The spectrum of $\hat{H}_{cp}$ contains four levels. The two higher levels are omitted, which corresponds to a merger of the two strongly coupled sites into a spin cluster in the presence of a (renormalized) transverse field~$\tilde{\Gamma}$ and a longitudinal field $\tilde{h}$. The magnetic moment of the cluster is given by $\tilde{\mu}=\mu_i+\mu_{i+1}$, where the
initial  magnetic moments are $\mu_i=\mu_{i+1}=1$. The remaining two lowest energy levels are given by second-order degenerate perturbation method as the eigenvalues of the matrix:
\begin{equation}
\begin{bmatrix}
 -J_i-h_i-h_{i+1}       & -\Gamma_i\Gamma_{i+1}/J_i \\
 -\Gamma_i\Gamma_{i+1}/J_i &  -J_i+h_i+h_{i+1} 
\end{bmatrix}
\end{equation}
having a gap 
\be
\Delta E_{cp}=2 \sqrt{\left(\Gamma_i\Gamma_{i+1}/J_i\right)^2+\left(h_i+h_{i+1}\right)^2}.
\ee
Comparing it with the gap of the spin cluster, $2\tilde{\gamma}$ in equation~(\ref{epsilon_i}), we obtain for the renormalised values of the parameters:
\be
\tilde{\Gamma}=\frac{\Gamma_i\Gamma_{i+1}}{J_i},\quad \tilde{h}=h_i+h_{i+1}.
\label{tilde_Gamma_h}
\ee

\subsubsection{Strong field decimation}
\label{sec:strong_field}

In this case, the largest local term is associated to a site, say $i$ as illustrated in figure~\ref{fig_0}, and the corresponding parameter of the one-site Hamiltonian is $\gamma_i$. Due to the large gap this site does not contribute to the longitudinal magnetisation and therefore it is eliminated. However, the longitudinal magnetic field should be transformed at the remaining neighbouring sites. To calculate the new renormalized coupling between the remaining sites $i-1$ and $i+1$, we calculate the energy levels with fixed spins at these sites. Denoting by $s_{i\pm1}=+$($-$) and $\uparrow$($\downarrow$) the boundary state, the eigenvalue problem with different boundary conditions has the lowest energy:
\begin{equation}
E_{s_{i-1},s_{i+1}}=-\sqrt{\Gamma_i^2+\left(s_{i-1}J_{i-1}+s_{i+1}J_i+h_i\right)^2}.
\end{equation}
The renormalised coupling between the remaining sites is given by:
\begin{equation}
\tilde{J}=-\left(E_{\uparrow \uparrow}+E_{\downarrow \downarrow}-E_{\uparrow \downarrow}-E_{\downarrow \uparrow}\right)/4 \approx \frac{J_{i-1}J_i \Gamma_i^2}{\left(\Gamma_i^2+h_i^2\right)^{3/2}},
\label{tilde_J}
\end{equation}
where the last relation is calculated perturbatively.

We obtain similarly for the excess longitudinal fields:
\be
\Delta h_{i-1}=-\left(E_{\uparrow \uparrow}-E_{\downarrow \downarrow}+E_{\uparrow \downarrow}-E_{\downarrow \uparrow}\right)/4 \approx \frac{J_{i-1}h_i}{\sqrt{\Gamma_i^2+h_i^2}},
\ee
and
\be
\Delta h_{i+1}=-\left(E_{\uparrow \uparrow}-E_{\downarrow \downarrow}-E_{\uparrow \downarrow}+E_{\downarrow \uparrow}\right)/4 \approx \frac{J_{i}h_i}{\sqrt{\Gamma_i^2+h_i^2}},
\ee
so that 
\be
\tilde{h}_{i\pm1}={h}_{i\pm1}+\Delta {h}_{i\pm1}.
\ee
Note that in our numerical implementation, we use the direct expressions with the energies, instead of the simpler perturbative approximation.
We also note that setting the longitudinal field $h_i=0$, the decimation equations in equations~(\ref{tilde_Gamma_h}) and (\ref{tilde_J}) are in the standard form, see in references~\cite{IGLOI2005277,Igloi2018}.

\subsection{Infinite disorder fixed point at $h_0=0$}
\label{sec:IDFP}
The properties of the system in the absence of random longitudinal fields were studied previously and several asymptotically exact results are known through the solution of the SDRG equations \cite{PhysRevLett.69.534,PhysRevB.51.6411}. 

The control parameter is defined as:
\be
 \delta=\frac{\overline{\ln \Gamma}-\overline{\ln J}}{\text{var}(\ln h)+\text{var}(\ln J)},    
 \label{eq:delta}
\ee
where $\overline{x}$ and $\text{var}(x)$ stands for the average and the variance of the random variable $x$, respectively. The critical point is at $\delta=0$, which corresponds to $\Gamma_0=1$ in the distribution in equation~(\ref{eq:J_distrib}). The energy scale, $\epsilon$, which is the smallest gap, scales with the length $L$ at the critical point as:
\be
\ln \epsilon \sim L^{\psi},
\label{psi}
\ee
with a critical exponent
\be
\psi=1/2.
\label{psi=1/2}
\ee
The magnetization moment, $\mu$ has also a power-law $L$-dependence at the critical point:
\be
\mu \sim L^{d_f},
\label{d_f}
\ee
where the fractal dimension of the magnetization is given by: 
\be
d_f=\big(1+\sqrt{5}\big)/4,
\label{d_f1}
\ee
which is just half of the golden ratio.

In the disordered phase $\delta>0$, the average correlations 
decay exponentially with the true correlation length 
$\xi \sim 1/\delta^2$, implying the correlation length exponent
\be
\nu=2\;
\label{nu}
\ee
for the random chain. On the other hand, the decay of the typical correlations involves a different exponent:
\be
\nu_{\rm typ}=1.
\label{nu_typ}
\ee
Next to the critical point in the disordered phase, in the so-called Griffiths phase, the energy-scale tends to zero as:
\be
\epsilon \sim L^{-z},
\label{L_z}
\ee
which results in singularities in the dynamical quantities, such as in the magnetic susceptibility or in the autocorrelation function. In equation~(\ref{L_z}),
$z$ is the dynamical exponent, which is given by the positive root of the equation \cite{PhysRevE.58.4238,PhysRevLett.86.1343,PhysRevB.65.064416}:
\be
      \overline{\left(\frac{J}{\Gamma} \right)^{1/z}}=1,
      \label{eq:z_eq}
\ee
which is an exact expression in the entire Griffiths region.
In the vicinity of  the critical point, the dynamical exponent to the leading order is given by \cite{grf}
\be
       z\approx \frac{1}{2|\delta|},\qquad |\delta| \ll 1.
\ee

\section{Numerical results for $h_0>0$}
\label{sec:results}

\subsection{Properties of the RG flow}
\label{sec:RG_flow}

\begin{figure}[h!]
	\centerline{\includegraphics[width=.6\columnwidth]{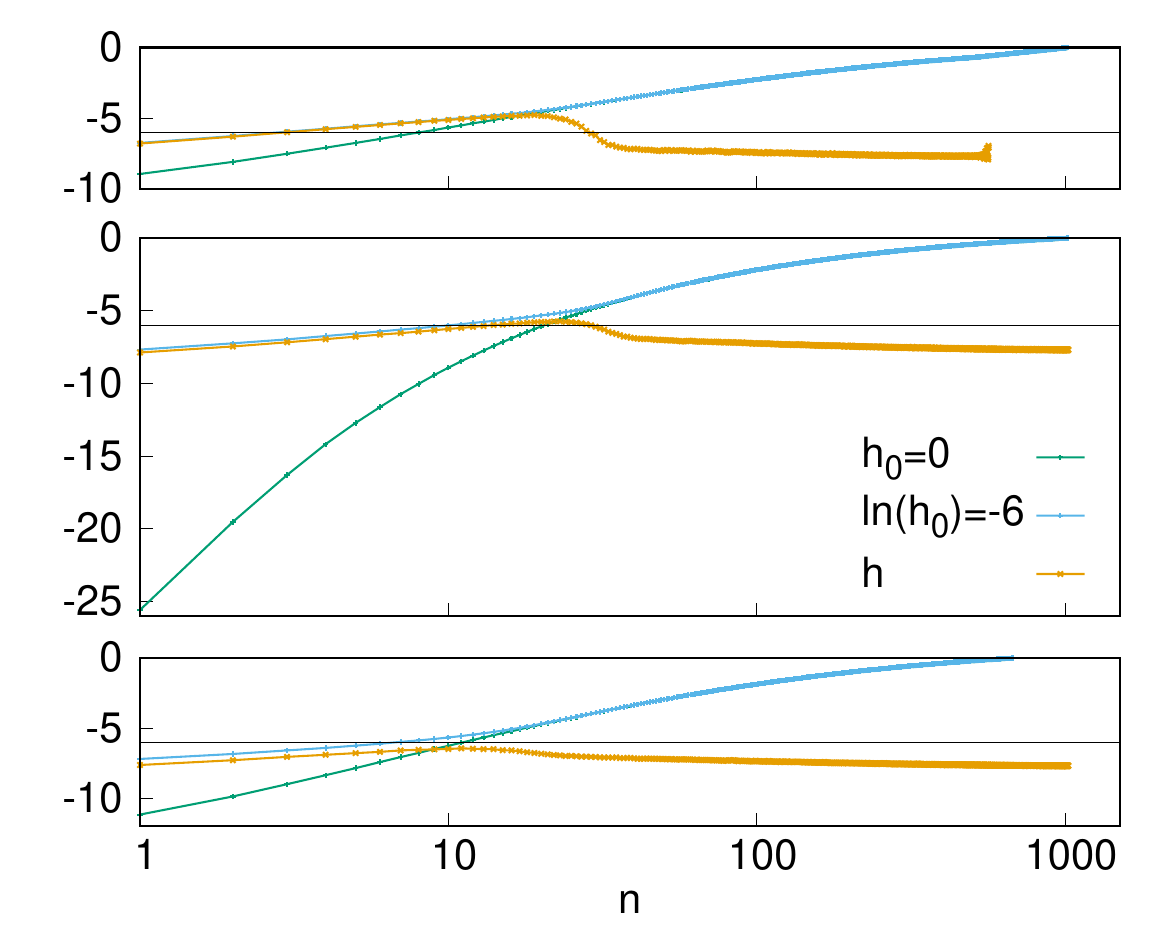}}
	\vskip -0.5cm
	\caption{(Colour online) Renormalized value of the log-energy excitations as a function of the remaining sites, $n$, for the non-perturbed model with $h_0=0$ (green symbols) and for the model with a random longitudinal field ($\ln h_0=-6$, blue symbols). The value of the log random field at a site decimation is shown by orange symbols. The results are obtained on a chain with $L=1024$ and the log-variables are averaged over $10\,000$ random samples. Upper panel: starting from the ordered unperturbed phase $\Gamma_0=0.5$; middle panel: starting from the unperturbed critical point $\Gamma_0=1.0$; lower panel:  starting from the disordered unperturbed phase $\Gamma_0=1.5$. The horizontal line at $\ln h_0=-6$ shows the parameter of the original distribution of the random longitudinal fields.}
	\label{fig_1a}	
\end{figure}

In this section, we switch on the random longitudinal fields and study their effect on the renormalization flow.
First, we study the behavior of the logarithmic excitation energy as a function of the remaining sites, $n$, and compare its behavior in the unperturbed system with $h_0=0$ to the behavior with finite perturbation $\ln h_0=-6$. 
The results for a chain of length $L=1024$ and averaged over $10\,000$ samples are shown in figure~\ref{fig_1a} at three different points: $\Gamma_0=0.5$ (starting from the ordered unperturbed phase), $\Gamma_0=1.0$ (critical point) and $\Gamma_0=1.5$ (disordered unperturbed phase). In these figures, we also monitor the renormalized value of the logarithmic longitudinal fields, which are involved in a site decimation. 

%%%%%%%%%%%%%%%%%%%%%%%%% Fig 1 %%%%%%%%%%
%%%%%%%%%%%%%%%%%%%%%

We can identify an \textit{initial period of the renormalization}, in which the excitation energies look similar for the two models, which is due to the fact that the renormalized longitudinal fields are comparatively small and do not influence the decimation rules. This behaviour is changed when the value of the renormalized longitudinal fields becomes comparable with the excitation energy. % and the system in the concluding RG steps renormalizes to the \textit{final state}. 
If we start from the ordered unperturbed phase, then in the initial period the couplings are dominantly decimated and the transverse  fields will be renormalized to an exponentially small value. In the second part of the renormalization, in the concluding RG steps (which means small values of $n$) the random longitudinal fields play a dominant role and the system finally renormalizes to a classical random-field Ising chain. On the contrary, if we start from the disordered phase, then in the initial period the fields are dominantly decimated and the couplings will be renormalized to an exponentially small value. In the concluding RG steps, the renormalization is influenced but not dominated by the random longitudinal fields and the system approaches a quantum disordered phase. Finally, if we start from the critical point of the unperturbed system, then in the initial period the couplings and fields are decimated in a symmetric way. In the concluding RG steps, the grown-up longitudinal fields will stop a further rapid decrease of the log excitation energy and the final state will be the result of all three parameters in the Hamiltonian.

\begin{figure}[h!]
\centerline{\includegraphics[width=.6\columnwidth]{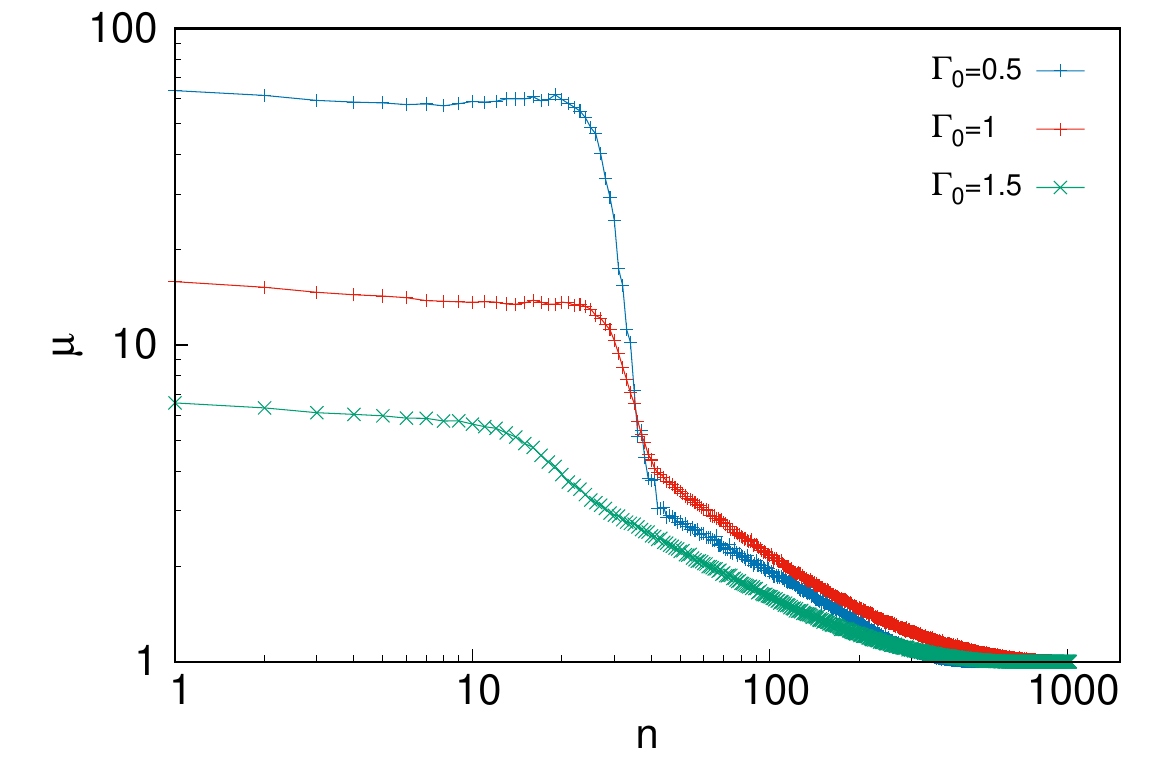}}
	\vskip -0.5cm
\caption{(Colour online) Evolution of the average cluster moment as a function of the remaining sites, $n$, for the model with a random longitudinal field ($\ln h_0=-6$) and for three values of the initial transverse field distributions: $\Gamma_0=0.5,1.0$ and $1.5$. The calculation details correspond to those in figure~\ref{fig_1a}.}
\label{fig_1aa}	
\end{figure}

The evolution of the average cluster moment during renormalization is shown in figure~\ref{fig_1aa} for three values of the initial transverse field distributions. As seen in this figure, in the initial period there is a slow increase of $\overline{\mu}$, which will turn to a fast increase when the random longitudinal fields also show an increase. In the final state, where the couplings are very rarely decimated, $\overline{\mu}$ shows a plateau, the height of which is larger for smaller values of $\Gamma_0$, as the remaining clusters stop growing and become gradually eliminated.

%%%%%%%%%%%%%%%%%%%%%%%%% Fig 2 %%%%%%%%%%
%%%%%%%%%%%%%%%%%%%%%

\subsection{RG phase-diagram}

%%%%%%%%%%%%%%%%%%%%%%%%% Fig 3 %%%%%%%%%%
\begin{figure}[h!]
\centerline{\includegraphics[width=.55 \columnwidth]{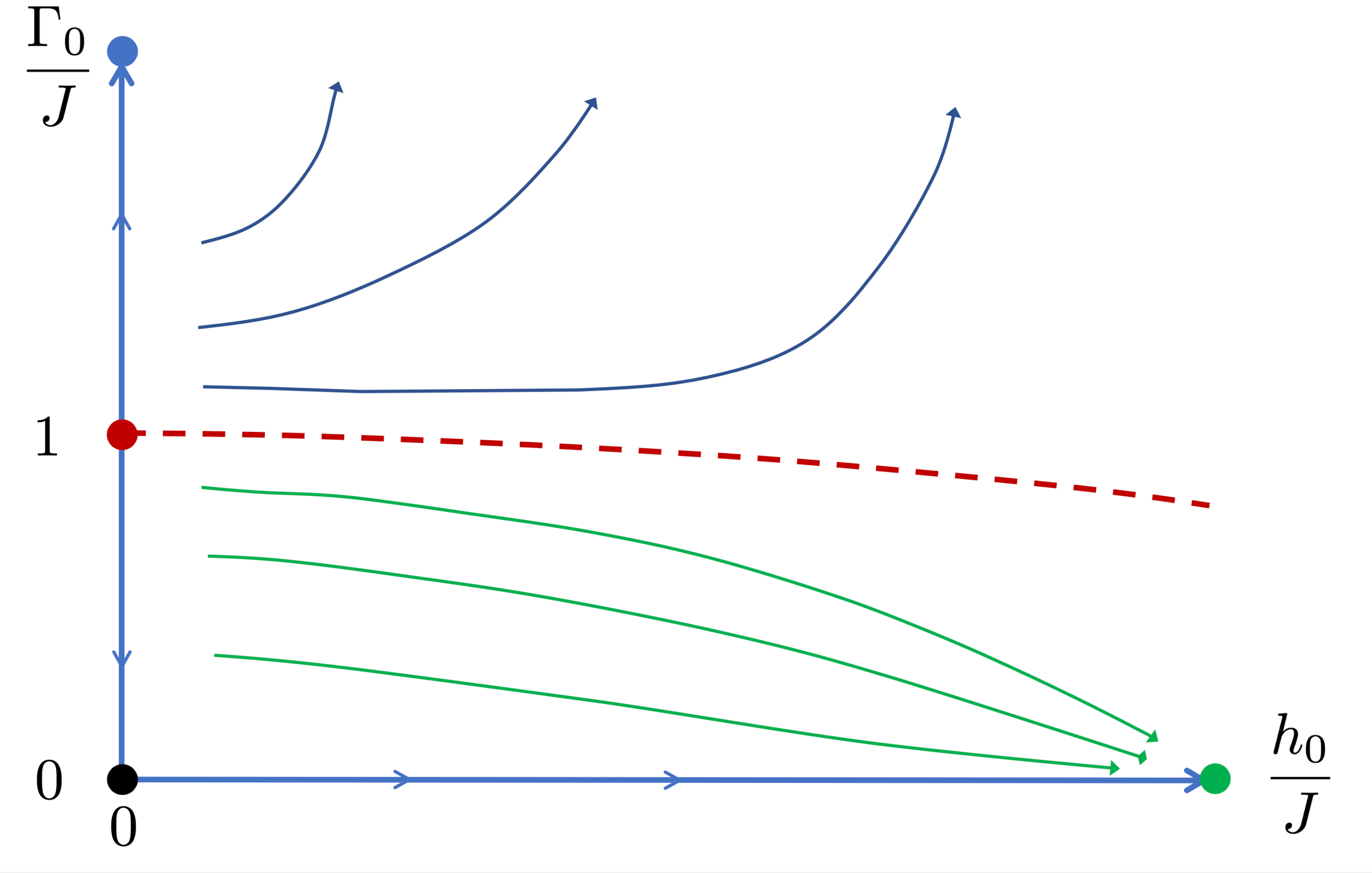}}
	\vskip -0.3cm
\caption{(Colour online) Schematic RG phase-diagram in the $L \to \infty$ limit. In the absence of random longitudinal fields at $h_0/J=0$, the IDFP (denoted by red circle) separates the quantum ordered phase ($\Gamma_0/J<1$) from the quantum disordered phase ($\Gamma_0/J>1$). At finite random longitudinal fields, there is a separatrix, indicated by a dashed red line, which starts from the IDFP and bends downwards. Below the separatrix, the RG flows (indicated by green lines) are attracted by the fixed point of the classical random field Ising model, which is shown by a green circle. Above the separatrix, the RG flows (indicated by blue lines) scale towards the quantum disordered phase.}
\label{fig_1b}	
\end{figure} 
%%%%%%%%%%%%%%%%%%%%%

Based on the previously observed properties of the RG flow, we suggest an RG phase-diagram which is shown in figure~\ref{fig_1b}.
In the absence of a random longitudinal field, at $h_0/J=0$, the phase-diagram is described in section~\ref{sec:IDFP}. It consists of an ordered phase, with a fixed point at $\Gamma_0/J=0$ and is indicated by a black circle in figure~\ref{fig_1b} and a quantum disordered phase, with a fixed point at $\Gamma_0/J \to \infty$, illustrated by a blue circle. The IDFP at $\Gamma_0/J=1$, which is shown by a red circle, separates the two phases. Switching on the random longitudinal field, the ordered phase will disappear. In the classical limit, at the line $\Gamma_0/J=0$, the properties of the system are controlled by the fixed point of the classical random field Ising model, which is located at $h_0/J \to \infty$ and illustrated by a green circle. For general values of the parameters, the system has two disordered phases, which are separated by a separatrix indicated by a dashed red line, which starts from the IDFP and bends downwards, due to the fact that the gap increases with increasing $h_0$, see in equation~(\ref{epsilon_i}). Below the separatrix, the RG flows are attracted by the fixed point of the classical random field Ising model, while above the separatrix, the RG flows scale towards the quantum disordered phase.

Here, we are going to study the properties of the system along the separatrix in figure~\ref{fig_1b} more in detail.

\subsection{Scaling behaviour in the vicinity of the IDFP}
\label{sec:Num_results}

Here, we perform a detailed numerical analysis of the renormalization results on finite periodic chains with lengths $L=2^k$, $k=7,8,\dots,14$, and for each length $100\,000$ random samples are considered. To treat such large samples, we used an efficient numerical algorithm which works in linear time as a function of $L$, up to a logarithmic correction.  

%%%%%%%%%%%%%%%%%%%%%%%%% Fig 4 %%%%%%%%%%
\begin{figure}[h!]
\includegraphics[width=.5\columnwidth]{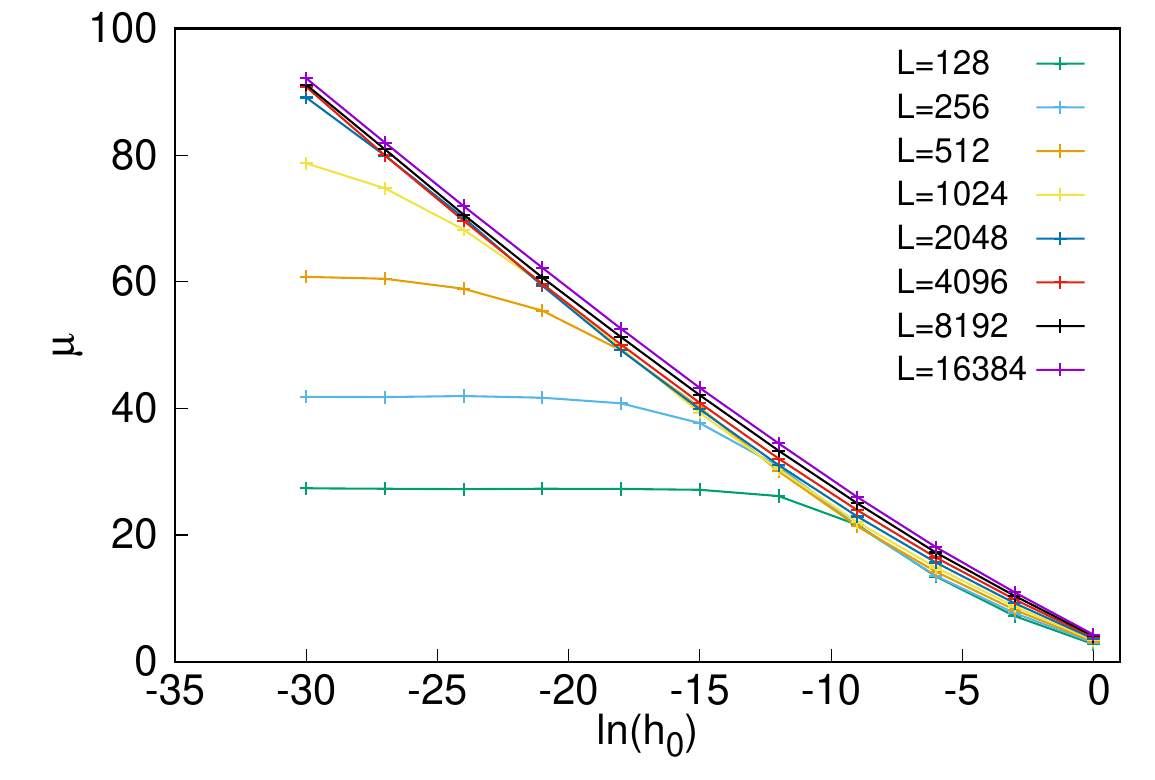}
\includegraphics[width=.5\columnwidth]{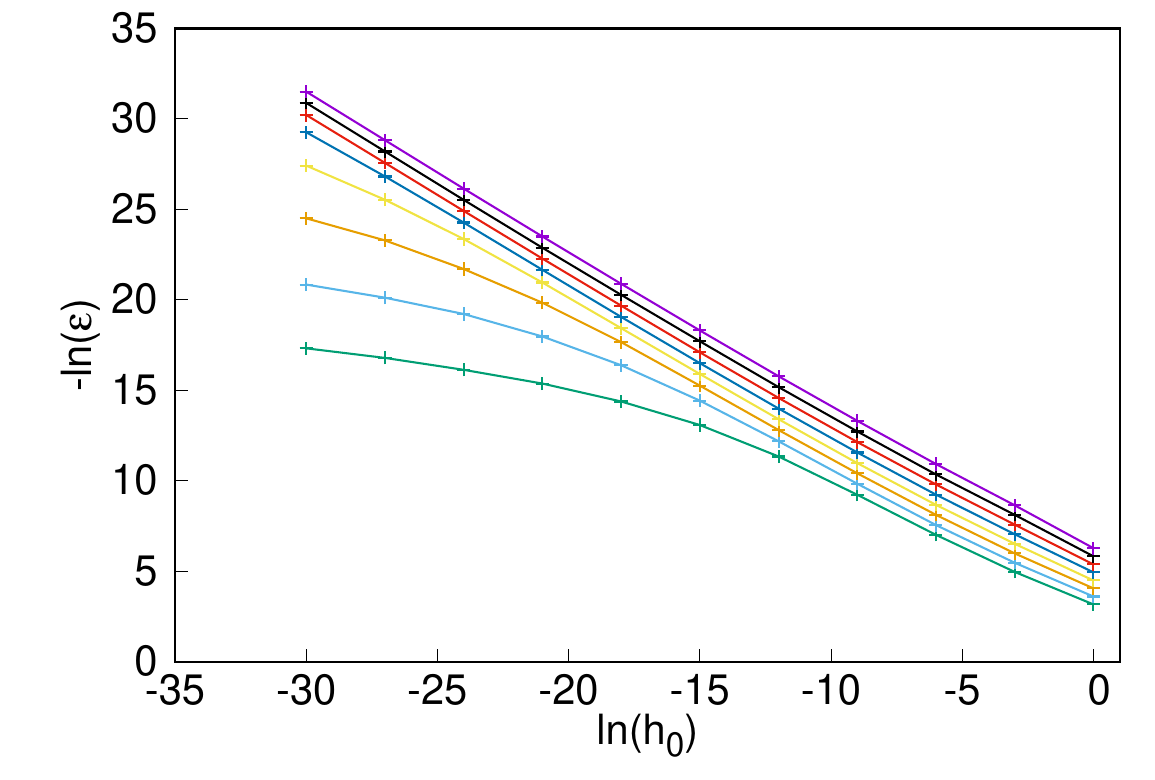}
	\vskip -0.4cm
\caption{(Colour online) Average magnetization moment (left-hand panel) and average log-gap (right-hand panel) as a function of $\ln h_0$ for different lengths of the chain.}
\label{fig_1}	
\end{figure} 
%%%%%%%%%%%%%%%%%%%%%

Having in mind the observation in the previous subsection, we consider the critical point of the system at $h_0=0$, which corresponds to $\Gamma_0=1$, and monitor the behaviour of the system at small values of $h_0$. It turned out that the appropriate variable is $\ln h_0$ and we made the calculations at $-\ln h_0=0, 3, 6,\dots,30$. We measured the average value of the magnetization moment, $\overline{\mu}$, as well as the average parameter of the log-gap, $\overline{\ln \epsilon}$, where $\epsilon$ is given by the last decimated site value: $\epsilon=\tilde{\gamma}=\sqrt{\tilde{\Gamma}^2+\tilde{h}^2}$. 

\begin{figure}[t!]
\includegraphics[width=.5\columnwidth]{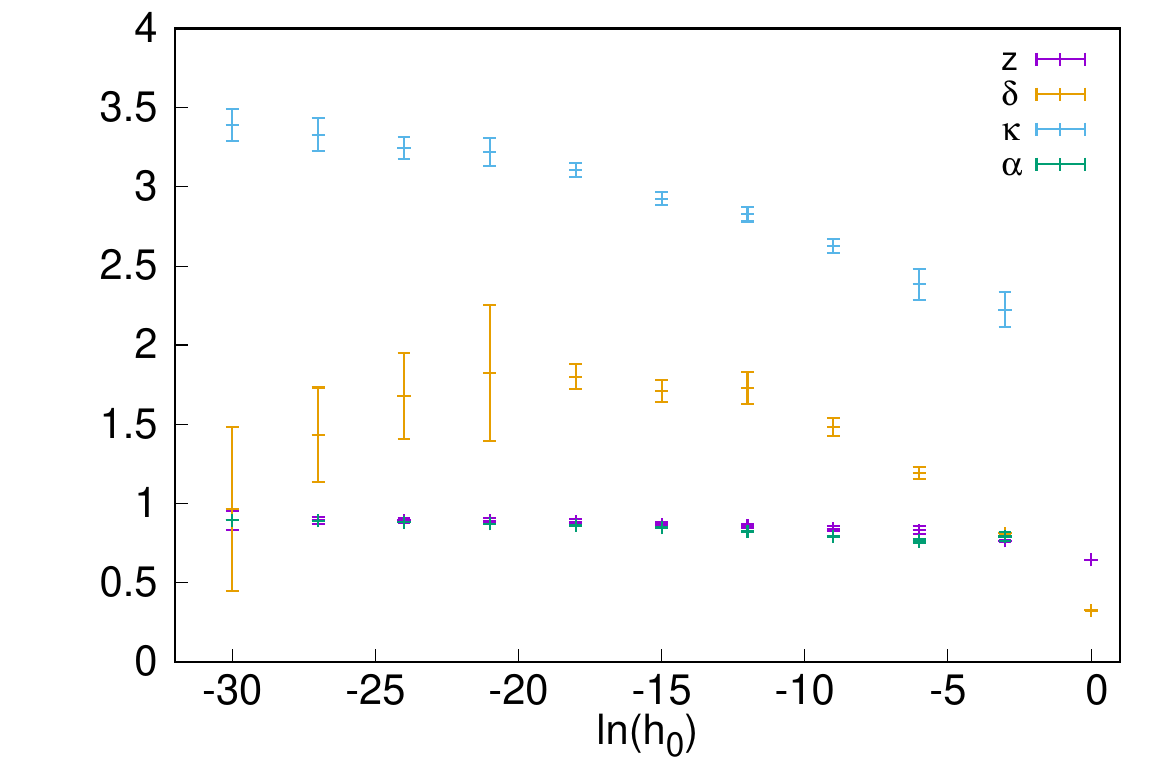}
\includegraphics[width=.5\columnwidth]{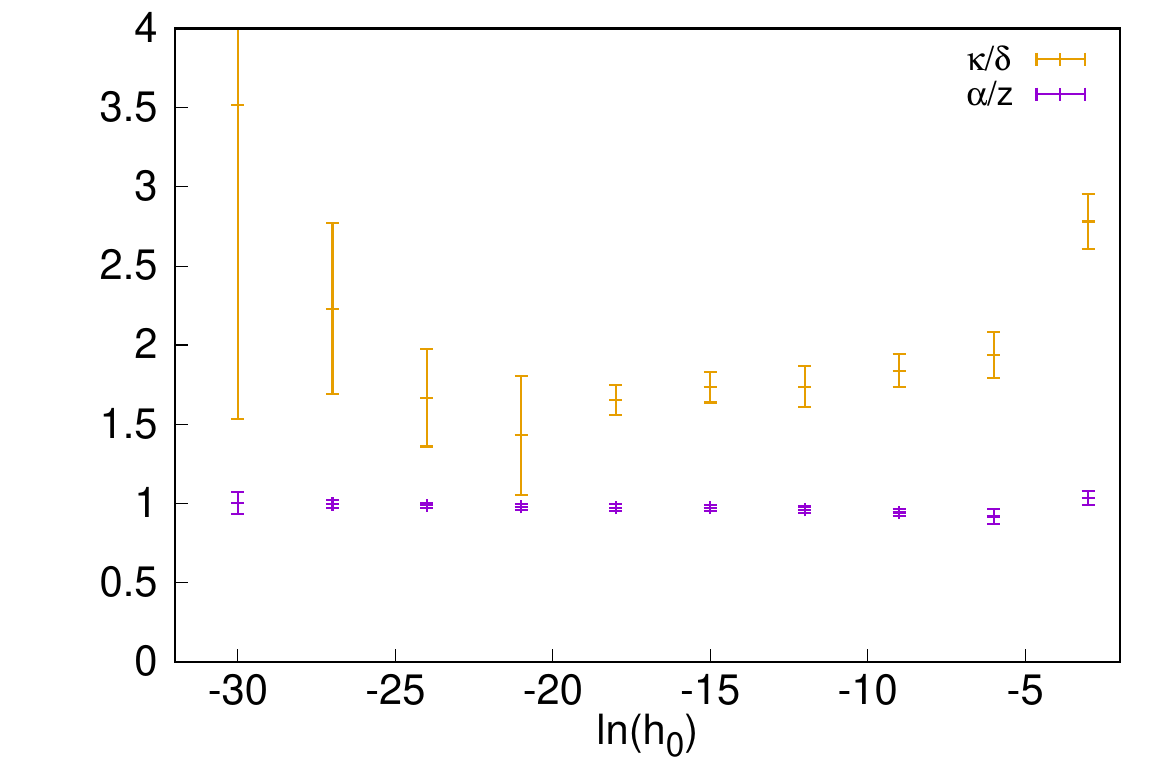}
	\vskip -0.4cm
\caption{(Colour online) Left-hand panel: estimated values of the local exponents in equations~(\ref{fix_L}) and~(\ref{fix_h_0}) for different parameters of the distribution of the random longitudinal field, $h_0$. Right-hand panel: ratio of the local exponents $\kappa/\delta$ and $\alpha/z$ for different values of $\ln h_0$.}
\label{fig_5}	
\end{figure}

According to the numerical results (see in figure~\ref{fig_1}), large sizes are needed to observe the final state, meaning 
% is observable for relatively large values of the length of the chain, which means 
that a large enough number of RG steps are performed. Equivalently, for a fixed chain length, $L$, the random longitudinal field should be sufficiently large, $h_0>\tilde{h}_0(L)$. For relatively small values of the longitudinal fields, $h_0<\tilde{h}_0(L)$ the influence of the original fixed point with $h_0=0$ is still strong and the system stays in the initial phase. Deep in the disordered phase, the average quantities have an approximately linear dependence on $\ln h_0$. Thus, between two close points, $h_0^{(2)}$ and $h_0^{(1)}$, we have the relations:
\begin{align}
\overline{\mu}_L\left(h_0^{(2)}\right)-\overline{\mu}_L\left(h_0^{(1)}\right) &\approx -\kappa \ln\left[h_0^{(2)}/h_0^{(1)}\right],\nonumber \\
\overline{\ln \epsilon}_L\left[h_0^{(2)}\right]-\overline{\ln \epsilon}_L\left[h_0^{(1)}\right] &\approx \alpha \ln\left[h_0^{(2)}/h_0^{(1)}\right].
\label{fix_L}
\end{align}
The prefactors, $-\kappa$ and $\alpha$, have a weak $\ln h_0$ dependence, as shown in the estimated values in figure~\ref{fig_5}.

%%%%%%%%%%%%%%%%%%%%%%%%% Fig 5 %%%%%%%%%%
\begin{figure}[t!]
\centerline{\includegraphics[width=.5\columnwidth]{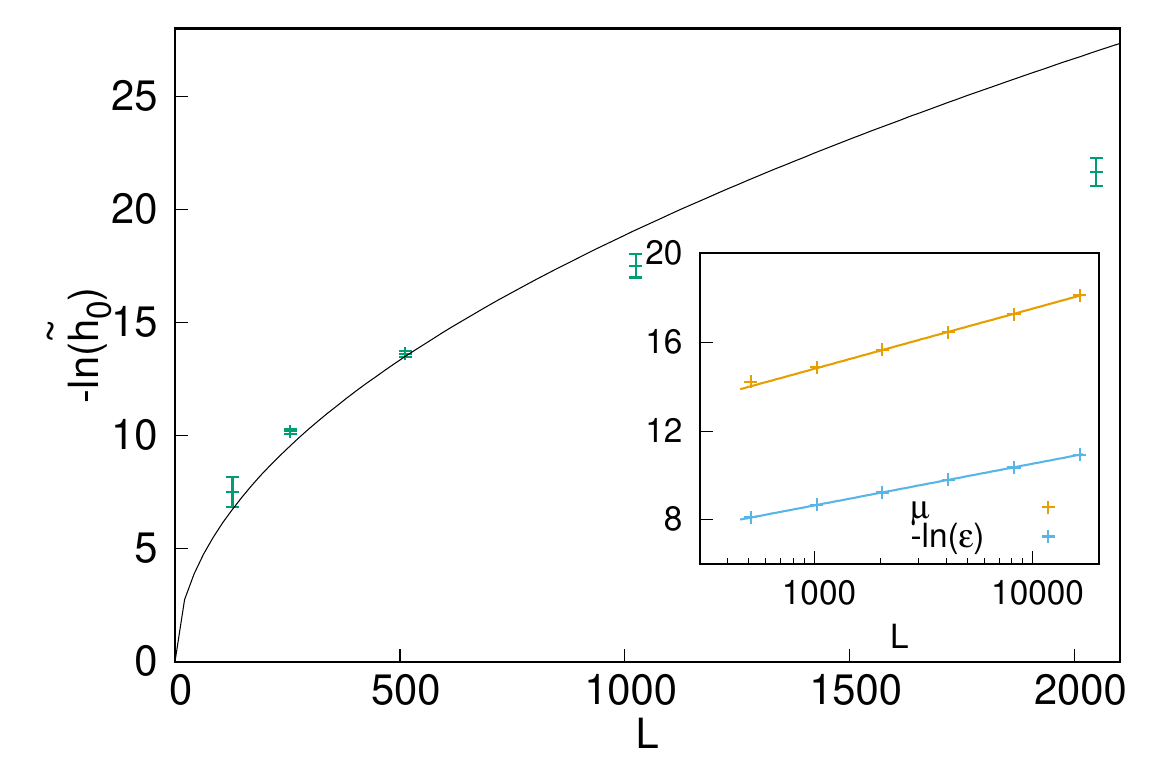}}
	\vskip -0.4cm
\caption{(Colour online) The boundary of the disordered phase as signalled by an inflexion point in the $\overline{\mu}$ and $\overline{\ln \epsilon}$ vs $\ln h_0$ curves, see in equation~(\ref{inflexion}). The full line illustrates an $\sim L^{1/2}$ dependence. 
Inset: size dependence of the average magnetization moment and the average log-gap at a fixed value of the longitudinal field $\ln h_0=-6$. The straight lines have the slopes $\delta=1.2$ and $z=0.817$, see in equation~(\ref{fix_h_0}).
}
\label{fig_2a}	
\end{figure}

%%%%%%%%%%%%%%%%%%%%%
This asymptotic regime ends at a limiting point, $\tilde{h}_0(L)$, which is approximately signalled by an inflection point, so that:
\be
\left.\frac{{\partial}^2\overline{\mu}_L}{{\partial}\ln h_0^2}\right|_{\tilde{h}_0(L)} \approx \left.\frac{{\partial}^2\overline{\ln \epsilon}_L}{{\partial}\ln h_0^2}\right|_{\tilde{h}_0(L)} \approx0.
\label{inflexion}
\ee
The positions of these inflection points are shown in figure~\ref{fig_2a}.
%%%%%%%%%%%%%%%%%%%%%%%%% Fig 6 %%%%%%%%%%
%%%%%%%%%%%%%%%%%%%%%
We notice an approximate relation:
\be
|\ln \tilde{h}_0(L)| < L^{1/2},
\ee
which can be explained as follows. At the IDFP (with $h_0=0$), the value of the typical excitation energy~$\epsilon_0$ scales as $|\ln \epsilon_0| \sim L^{1/2}$, see in equation~(\ref{psi=1/2}). If $|\ln \tilde{h}_0(L)| > |\ln \epsilon_0|$, then the perturbation caused by the random longitudinal field is much smaller than the smallest excitation energy. Consequently, it cannot influence the critical properties of the system. If we try to fit the measured values as $|\ln \tilde{h}_0(L)| \sim L^{\omega}$, we have $\omega<0.5$. The effective exponent is observed to decrease with $L$, c.f. at $L=2048$, we have $\omega_{\rm eff}\approx 0.3$. This is due to the fact that during renormalization the typical value of $\sqrt{\Gamma^2+h^2}$ will be larger than the typical value of $J$. Thus, the system moves to the disordered phase. Here, almost always strong field decimations take place (see in section~\ref{sec:strong_field}) and as a consequence $\ln \epsilon$ will have a weaker (presumably logarithmic) $L$ dependence.

If we consider the curves at a fixed value of $h_0$ for different values of $L$, these are merely shifted. Thus, for two lengths, $L_2$ and $L_1$ we have the relations:
\begin{align}
\overline{\mu}_{L_2}(h_0)-\overline{\mu}_{L_1}(h_0) &\approx \delta \ln (L_2/L_1),\nonumber \\
\overline{\ln \epsilon}_{L_2}(h_0) -\overline{\ln \epsilon}_{L_1}(h_0)&\approx - z \ln (L_2/L_1).
\label{fix_h_0}
\end{align}
This behaviour is illustrated in the inset of figure~\ref{fig_2a} and the estimated prefactors are shown in figure~\ref{fig_5}, which have also a weak $\ln h_0$ dependence.

Combining the relations in equations~(\ref{fix_L}) and (\ref{fix_h_0}), we obtain for the magnetic moment:
\be
\overline{\mu}_{L_{\mu}}(h_0) \approx \overline{\mu}_{L_1}\left(h_0^{(1)}\right) -\kappa \ln\left[h_0/h_0^{(1)}\right] + \delta \ln (L_{\mu}/L_1),
\ee
and similarly for the average log-gap:
\be
\overline{\ln \epsilon}_{L_{\epsilon}}(h_0) \approx \overline{\ln \epsilon}_{L_1}\left(h_0^{(1)}\right) +\alpha \ln\left[h_0/h_0^{(1)}\right] -z \ln (L_{\epsilon}/L_1).
\label{epsilon_rel}
\ee

Let us fix $L_1$ and $h_0^{(1)}$ at some reference value and keep the same value for the average magnetic moment, $\overline{\mu}_{L_{\mu}}(h_0) \approx \overline{\mu}_{L_1}\big(h_0^{(1)}\big)$. This is satisfied if we have a relation between the length associated to magnetic moments, $L_{\mu}$ and the distance from the fixed point, $h_0$ as:
\be
L_{\mu} \sim h_0^{-\nu_{\mu}},\quad \nu_{\mu}=\kappa/\delta.
\ee
Similar analysis of the expression for the log-gap in equation~(\ref{epsilon_rel}) leads to the relation:
\be
L_{\epsilon} \sim h_0^{-\nu_{\epsilon}},\quad \nu_{\epsilon}=\alpha/z,
\ee
where $L_{\epsilon}$ is the length associated to the energy gap.

Estimates for the correlation length exponents $\nu_{\mu}=\kappa/\delta$ and $\nu_{\epsilon}=\alpha/z$ are shown in the right-hand panel of figure~\ref{fig_5}. For small values of $h_0$, within the error of the approximation, these are in agreement with the values $\nu_{\mu} \approx 2$ and $\nu_{\epsilon} \approx 1$. These relations can be interpreted in the following way. Let us first consider the magnetization. At a given value of $h_0>0$, the sites after $L \to \infty$ RG steps are typically decimated out so that the average magnetization vanishes: $m=\lim_{L \to \infty}\overline{\mu}_L(h_0)/L=0$. After a finite number ($L < \infty$) of RG steps, however, effective spin clusters are formed having an average moment: $\overline{\mu}_L(h_0)$. The typical distance between such two clusters defines the length-scale $L_{\mu}(h_0)$ and, according to the previous measurement, $L_{\mu}^{1/\nu_{\mu}} h_0 \sim \tilde{h}={\cal O}(1)$. Consequently,  $\nu_{\mu}$ is the critical exponent of the (average magnetization) correlation length. We can estimate $\nu_{\mu}$ assuming that the increase of $\tilde{h}$ during renormalization is dominantly due to the process in equation~(\ref{tilde_Gamma_h}), which would mean the sum of $\sim L_{\mu}$ random numbers with the distribution in equation~(\ref{eq:J_distrib}). As a consequence, $\nu_{\mu}=2$ according to the central limit theorem. This result agrees with the correlation length exponent at $h_0=0$, see in equation~(\ref{nu}). The finite-size magnetization is related to average fluctuations of the parameters, while the smallest energy gap is due to extreme fluctuations of the parameters in a rare region. In such a rare region, the renormalized longitudinal field grows in an extreme way, which means a linear relation with its size. Thus, the $h_i$ parameters here have dominantly the same sign. The renormalized longitudinal fields start to enter in the decimation process, if at some site $\tilde{h}_i \sim  h_0 L_{\epsilon}={\cal O}(1)$. Consequently, the typical size of such rare region is $L_{\epsilon} \sim h_0^{-1}$, %With this assumption we arrive at the relation: $\tilde{h} \sim  h_0 L_{\epsilon}={\cal O}(1)$, 
thus $\nu_{\epsilon}=1$. This is just the critical exponent of the typical correlations in equation~(\ref{nu_typ}).

\section{Discussion}
\label{sec:disc}
In this paper, we considered the Ising model with random couplings and in the presence of simultaneous random transverse and longitudinal fields. This investigation is motivated by experimentally realized compounds, such as
${\rm LiHo}_x{\rm Y}_{1-x}{\rm F}_4$, which is placed into a magnetic field, transverse to the Ising axis. Here, we considered the one-dimensional system with nearest neighbour interaction and studied its low-energy properties by the strong disorder renormalization group approach. A similar model, although with a staggered non-random longitudinal field was studied recently \cite{PhysRevB.96.064427,PhysRevB.101.024203}, while other versions were studied in the context of many-body localization  \cite{MBL1,MBL2,MBL3,MBL4,MBL5,MBL6,MBL7}. 
 
In the absence of longitudinal field, the model exhibits an ordered and a quantum disordered phase, which are separated by an infinite disorder fixed point (IDFP). Switching on the random longitudinal field, the ordered phase disappears and the RG trajectories are attracted by two disordered phases, as shown in figure~\ref{fig_1b}. There is a separatrix, which is connected to the IDFP and below which the RG trajectories are attracted by the classical random field Ising chain fixed point. Above the separatrix, the trajectories approach a disordered phase, dominated by transverse-field effects.

We have studied the scaling properties of the system in the vicinity of the IDFP more in detail, since the SDRG is expected to be more accurate in this region. Approaching the IDFP on the separatrix, the correlation length, $\xi_{\mu}$, measured as the distance between two non-decimated magnetic moments, diverges as
$\xi_{\mu} \sim h_0^{\nu_{\mu}}$, with $\nu_{\mu} \approx 2$. On the other hand,  the distance between two non-decimated rare events of low-energy excitations, $\xi_{\epsilon}$ diverges with a different exponent: $\nu_{\epsilon} \approx 1$. Since the value of $\nu_{\mu}$ and $\nu_{\epsilon}$ corresponds to the exponents along the $h_0=0$ line for the average and typical correlation lengths, respectively, we expect that scaling at the IDFP is isotropic. We have also measured the value of the dynamical exponent,
which is found to be $h_0$ dependent and which approaches the value $z \approx 0.9$, as $h_0 \to 0$. Since $z(h_0=0)$ is formally infinity, this means that the dynamical exponent has a discontinuity at $h_0=0$. 

In our model, the distribution of the random longitudinal field is symmetric: $p(h)=p(-h)$. A non-symmetric distribution with a non-zero mean value $\overline{h}=\int h p(h)\, \rd h \ne 0$ has a different effect close to the IDFP. As shown previously \cite{PhysRevB.51.6411,PhysRevLett.69.534,IGLOI2005277}, longitudinal magnetization is induced and the corresponding susceptibility scales as
\be
\chi(\overline{h}) \sim \frac{\left|\ln \overline{h}\right|^{-x_m/\psi}}{\overline{h}},
\ee
with $x_m=1-d_f$, see the exponents in equations~(\ref{psi=1/2}) and (\ref{d_f1}).

%%%%%%%%%%%%%%%%%%%%%%%%% Fig 7 %%%%%%%%%%
\begin{figure}[h!]
\centerline{\includegraphics[width=.5 \columnwidth]{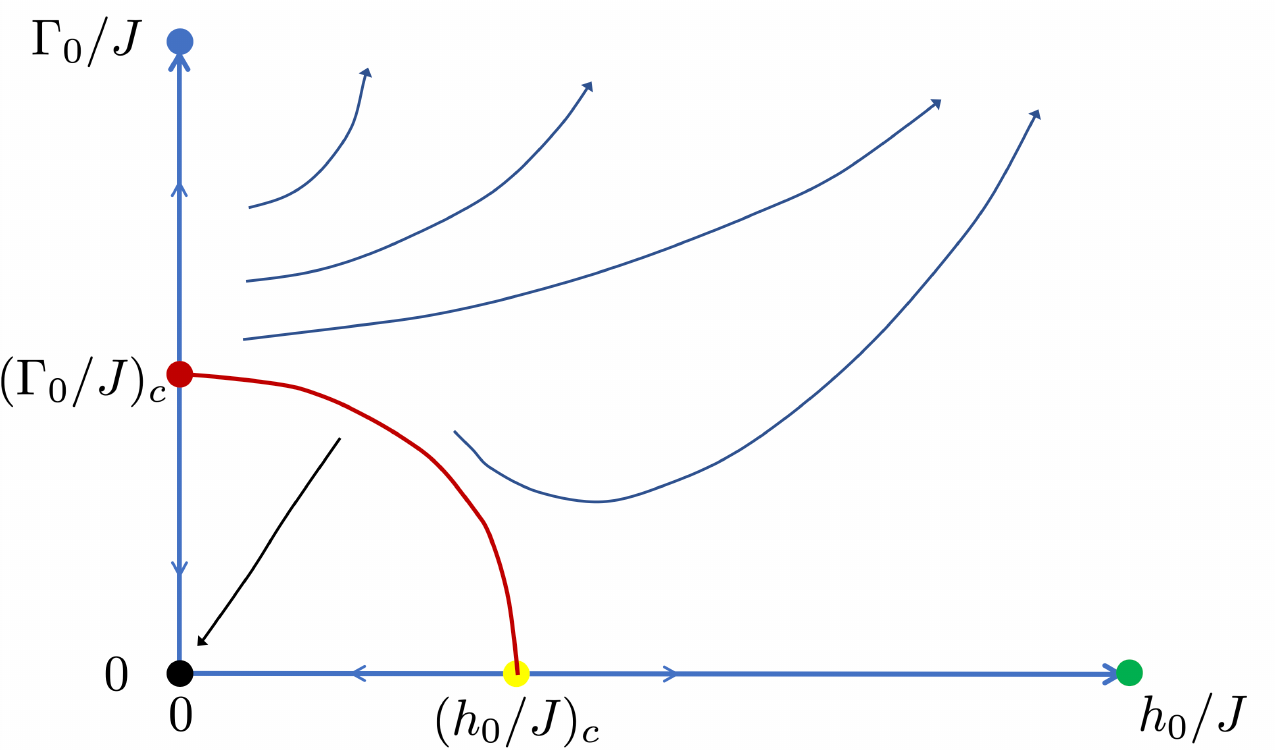}}
	\vskip -0.3cm
\caption{(Colour online) Expected schematic RG phase-diagram for the three-dimensional model with short-range interactions. In the absence of random longitudinal fields, at $h_0/J=0$, the IDFP (denoted by red circle) separates the quantum ordered phase, $\Gamma_0/J<(\Gamma_0/J)_c$, from the quantum disordered phase, $\Gamma_0/J>(\Gamma_0/J)_c$. Without transverse fields, at $\Gamma_0/J=0$, the ordered phase survives until $(h_0/J)_c$, indicated by a yellow circle. The red line connecting the red and yellow circles indicates the border between the ordered and the disordered phases. The RG flows, blue and black lines, are illustrations, their concrete form depends on the stability of the fixed points.}
\label{fig_7}	
\end{figure} 
%%%%%%%%%%%%%%%%%%%%%

One might ask the question about the possible modification of the RG phase-diagram in higher dimensions. In $d=2$, the phase-diagram is expected to be qualitatively the same as in figure~\ref{fig_1b}, since the random-field Ising model has no ordered phase \cite{Binder1983} in this case, either. In $d=3$, however, where the ordered phase survives for small values of $h_0/J<(h_0/J)_c$ \cite{PhysRevLett.35.1399,PhysRevLett.59.1829}, the phase-diagram should contain an ordered and a disordered phase. The expected phase-diagram is shown, see figure~\ref{fig_7}. Another problem to study is whether the random interaction is long-ranged, as c.f. in the compound ${\rm LiHo}_x{\rm Y}_{1-x}{\rm F}_4$, having dipole couplings. This problem in the absence of random longitudinal fields has a conventional random fixed point \cite{Juh_sz_2014,PhysRevB.93.184203} and further detailed studies are needed to obtain the phase-diagram for a complete model.

\section*{Acknowledgements}
%\begin{acknowledgments}
This paper is dedicated to Bertrand Berche on the occasion of his sixtieth birthday. One of us F.~I. has known Bertrand for more than thirty years and they have several works together. Bertrand's profound knowledge and interest in various areas of physics and his social commitment are greatly appreciated. 

This work was supported by the National Research Fund under Grants No. K128989 and No. KKP-126749, and by the National Research, Development and Innovation Office of Hungary (NKFIH) within the Quantum Information National Laboratory of Hungary.
%\end{acknowledgments}

\ukrainianpart

\title{Випадковий ланцюжок Ізінга у поперечному та поздовжньому полях: РГ-дослідження при сильному безладі}
\author{Т. Пето\refaddr{label1},
	Ф. Іглої\refaddr{label2,label1}, І. А. Ковач\refaddr{label3,label4}}

\addresses{
	\addr{label1} Інститут теоретичної фізики, Сегедський університет, H-6720 Сегед, Угорщина
	\addr{label2} Фізичний дослідницький центр ім. Вігнера, Інститут фізики та оптики твердого тіла, H-1525 Будапешт, 49, Угорщина
	\addr{label3} Факультет фізики та астрономії, Північно-Західний університет, Еванстон, Іллінойс 60208, США
	\addr{label4} Північно-Західний інститут складних систем, Північно-Західний університет, Еванстон, Іллінойс 60208, США}
\makeukrtitle 

\begin{abstract}
	\tolerance=3000%
	Маючи на увазі сполуку ${\rm LiHo}_x{\rm Y}_{1-x}{\rm F}_4$, розглянуто ланцюжок Ізінга з випадковими зв’язками та за наявності одночасних випадкових поперечних і поздовжніх полів, та вивчено його низькоенергетичні властивості при нульовій температурі за допомогою ренормгрупового підходу в умовах сильного безладу.	
	За відсутності поздовжніх полів у системі спостерігаються впорядкована або невпорядкована квантові фази, розділені критичною точкою безмежного безладу.
	Коли вмикається поздовжнє випадкове поле, впорядкована фаза зникає і траєкторії ренормгрупи притягуються до двох невпорядкованих фіксованих точок: одна з них характерна для класичного ланцюжка Ізінга у випадковому полі, інша описує квантову невпорядковану фазу. Дві невпорядковані фази розділені сепаратрисою, яка починається у фіксованій точці безмежного безладу, поблизу якої спостерігаються сильні квантові флуктуації.

	
\end{abstract}

\lastpage
\end{document}